\documentclass[a4paper]{spie}  
\usepackage[utf8x]{inputenc} 
\usepackage[]{graphicx}
\usepackage{subfigure}
\usepackage{transparent}
\usepackage{amssymb}
\usepackage{amsmath} 
\usepackage[english]{babel}
\usepackage{multicol}

\newcommand{\e}[1]{\times 10^{#1}} 
\newcommand{\ml}[1]{\mathrm{#1}} 
\newcommand{\dg}{^{\circ}}
\newcommand{\specialcell}[2][l]{%
  \begin{tabular}[#1]{@{}l@{}}#2\end{tabular}}

\title{Metrology calibration and very high accuracy centroiding with the NEAT testbed} 

\author{A. Crouzier\supit{a}, F. Malbet\supit{a}, O. Preis\supit{a}, F. Henault\supit{a}, P. Kern\supit{a}, G. Martin\supit{a}, P. Feautrier\supit{a}, E. Stadler\supit{a}, S. Lafrasse\supit{a}, A. Delboulbe\supit{a}, E. Behar\supit{a}, M. Saint-Pe\supit{a}, J. Dupont\supit{a}, S. Potin\supit{a}, C. Cara\supit{b}, M. Donati\supit{b}, E. Doumayrou\supit{b}, P. O. Lagage\supit{b}, A. Léger\supit{c}, J. M. LeDuigou\supit{d}, M. Shao\supit{e}, R. Goullioud\supit{e}
\skiplinehalf
\supit{a}Institut d'Astrophysique et de Planétologie de Grenoble, 414 Rue de la Piscine, St Martin d'Hères, Grenoble, France; \\
\supit{b}Commissariat à l'Energie Atomique et aux Energies Alternatives, Saclay, centre d'études nucléaires de Saclay, Paris, France; \\
\supit{c}Institut d'Astrophysique Spatiale, Centre universitaire d'Orsay, Paris, France; \\
\supit{d}Centre National d'Etudes Statiales, 2 place Maurice Quentin, Paris, France; \\
\supit{e}Jet Propulsion Laboratory, 4800 Oak Grove Drive, Pasadena, CA, U.S.A. 91109
}

\begin{document} 
\maketitle 
\graphicspath{ {./Figures/} }
\begin{abstract}

NEAT is an astrometric mission proposed to ESA with the objectives of detecting Earth-like exoplanets in the habitable zone of nearby solar-type stars. NEAT requires the capability to measure stellar centroids at the precision of $5\e{−6}$ pixel. Current state-of-the-art methods for centroid estimation have reached a precision of about $2\e{−5}$ pixel at two times Nyquist sampling, this was shown at the JPL by the VESTA experiment. A metrology system was used to calibrate intra and inter pixel quantum efficiency variations in order to correct pixelation errors. The European part of the NEAT consortium is building a testbed in vacuum in order to achieve $5\e{−6}$ pixel precision for the centroid estimation. The goal is to provide a proof of concept for the precision requirement of the NEAT spacecraft. 

The testbed consists of two main sub-systems. The first one produces pseudo stars: a blackbody source is fed into a large core fiber and lights-up a pinhole mask in the object plane, which is imaged by a mirror on the CCD. The second sub-system is the metrology, it projects young fringes on the CCD. The fringes are created by two single mode fibers facing the CCD and fixed on the mirror. In this paper we present the experiments conducted and the results obtained since July 2013 when we had the first light on both the metrology and pseudo stars. We explain the data reduction procedures we used.

\end{abstract}




\section{INTRODUCTION}\label{sec:INTRODUCTION} 

\subsection{Presentation of the Nearby Earth Astrometric Telescope (NEAT)}\label{subsec:Presentation of the Nerby Earth Astrometric Telescope}

With the present state of exoplanet detection techniques, none of the rocky planets of the Solar System would be detected. By measuring the reflex motion of planets on their central host stars, astrometry can yield the mass of planets and their orbits. However it is necessary to go to space to reach the precision required to detect all planets down to the Earth mass.
\\

\begin{figure}[t]
\begin{center}
\includegraphics[height = 60mm]{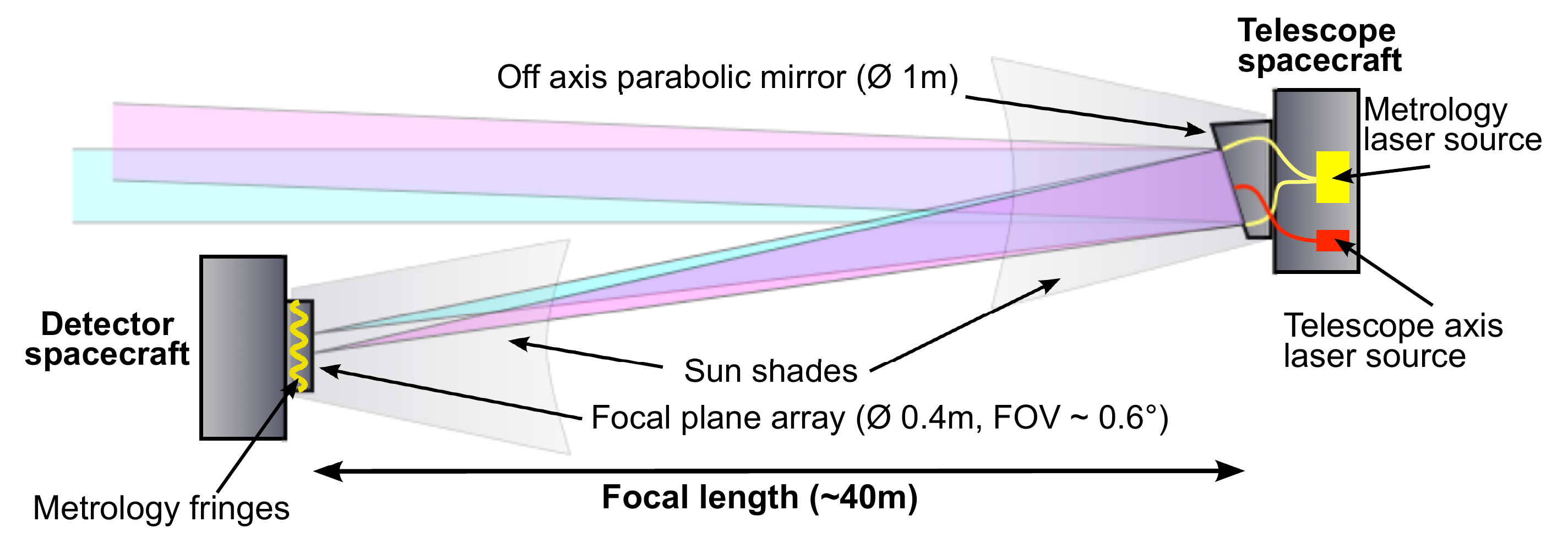}
\caption{\label{neat_concept_diagram}\textbf{The proposed NEAT concept.} The metrology system projects dynamic Young fringes on the detector plane. The fringes allow a very precise calibration of the CCD in order to reach micro-pixel centroiding errors.}
\end{center}
\end{figure}

The NEAT consortium is proposing a mission to ESA in the framework of the call for M missions in the Cosmic Vision plan which objective is to find most of the exoplanets of our Solar neighbourhood\cite{Malbet11,Malbet12,Malbet13,Malbet14}. The goal is to use differential astrometry to complete the measurements obtained by other techniques, thus lowering the threshold of detection and characterization down to the level of an Earth mass in the habitable zone of each system. We want to explore in a systematic manner all solar-type stars (FGK spectral type) up to 20 pc from the Sun. The satellite concept is based on formation flying with a satellite carrying a single primary mirror and another satellite carrying the focal plane (see Fig.~\ref{neat_concept_diagram}). The measurement is done using laser metrology and interferometry.

One of the fundamental aspects of the NEAT mission is the extremely high precision required to detect exo-Earths in habitable zone by astrometry. The amplitude of the astrometric signal that a planet leaves on its host star is given by the following formula:
\begin{equation}\label{eq:astrometric_signal}
A = 3 \mu \ml{as} \times \frac{M_{\ml{Planet}}}{M_{\ml{Earth}}} \times \left(\frac{M_{\ml{Star}}}{M_{\ml{Sun}}}\right)^{-1} \times \frac{R}{1\ml{AU}} \times \left(\frac{D}{1\ml{pc}}\right)^{-1}
\end{equation}

Where $D$ is the distance between the sun and the observed star, $M_{\ml{Planet}}$ is the exoplanet mass, $R$ is the exoplanet semi major axis and $M_{\ml{Star}}$ is the mass of the observed host star. For an Earth in the habitable zone located at 10 pc from the sun, the astrometric signal is 0.3 micro arcseconds (or $1.45\e{-11}$ rad). With a focal length of 40 meters, and taking into account a required signal to noise ratio\cite{sim_double_blind_test09} of 6 and the required number of measurements per target \cite{neat_number_of_measurements11}, the 0.3 $\mu$as requirement corresponds to a calibration of pixelation errors to $5\e{-6}$, as shown by the NEAT error budget\cite{neat_error_budget11}.

\subsection{Presentation of the NEAT testbed}\label{subsec:Presentation of the NEAT testbed}

\begin{figure}[t]
\begin{center}
\includegraphics[height = 80mm]{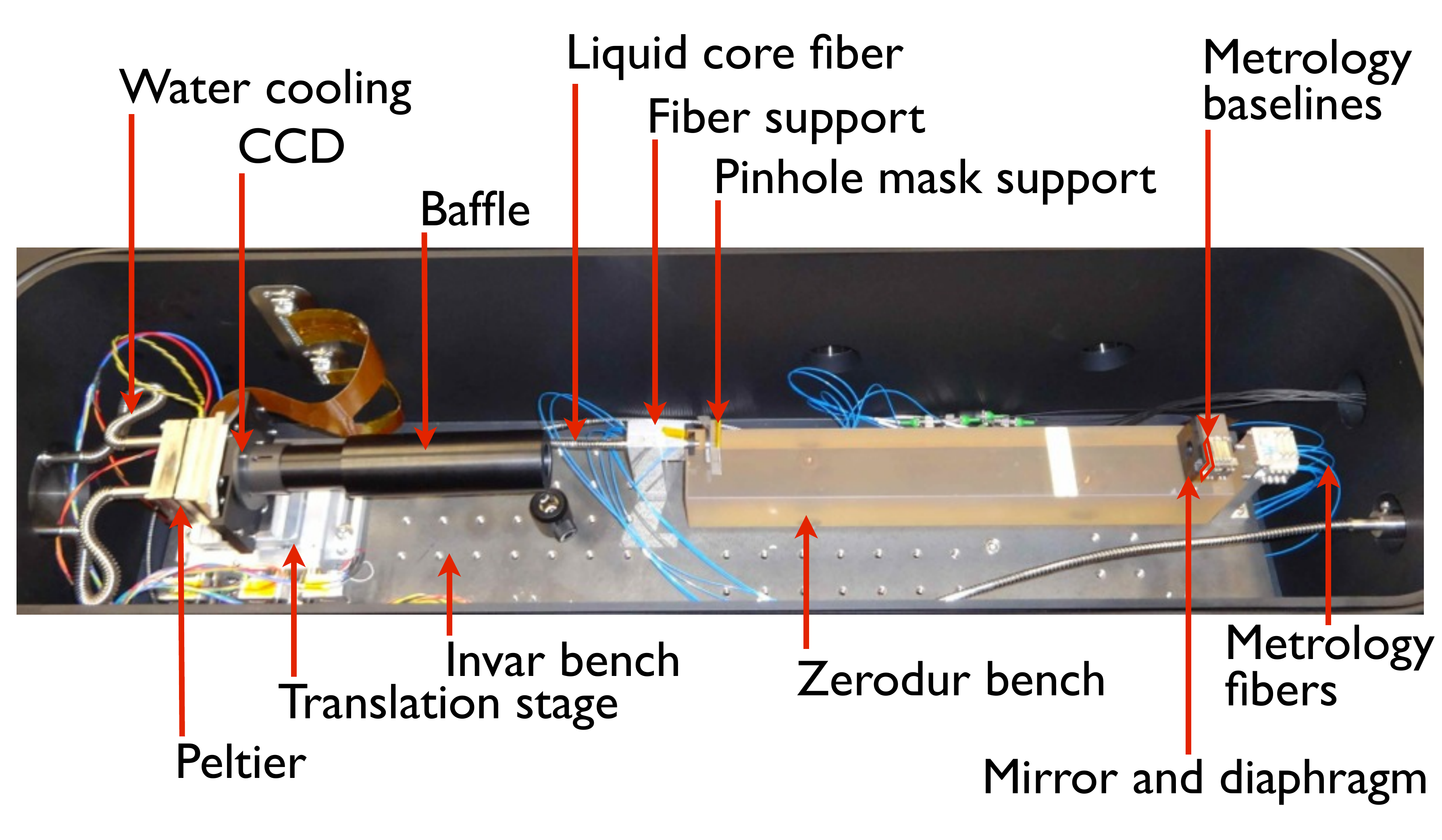}
\caption{\label{schematic_vacuum_chamber}\textbf{Schematic of the optical bench of the NEAT testbed.} The part shown here is the bench inside the vacuum chamber, which mimics the optical configuration of the NEAT spacecraft: it is the core of the experiment. Two main peripheral sub-systems, the metrology and pseudo stellar sources produce respectively fringes and stars on the CCD.}
\end{center}
\end{figure}

In order to demonstrate the feasibility of the calibration, the NEAT testbed has been assembled at IPAG. Figure \ref{schematic_vacuum_chamber} is a labeled picture of the inside of the vacuum chamber. Here we will not elaborate any longer about the testbed, in previous SPIE papers we have already presented: the testbed itself, the context around it, its specifications, a photometric budget (Amsterdam, 2012)\cite{Crouzier12}{}, the results after the first light was obtained in July 2013 and an error budget (San Diego, 2013)\cite{Crouzier13}{}. A last paper presents a more complete and updated error budget (Montreal, 2014)\cite{Henault14}{}. 

\begin{figure}[t]
\begin{center}
\includegraphics[height = 80mm]{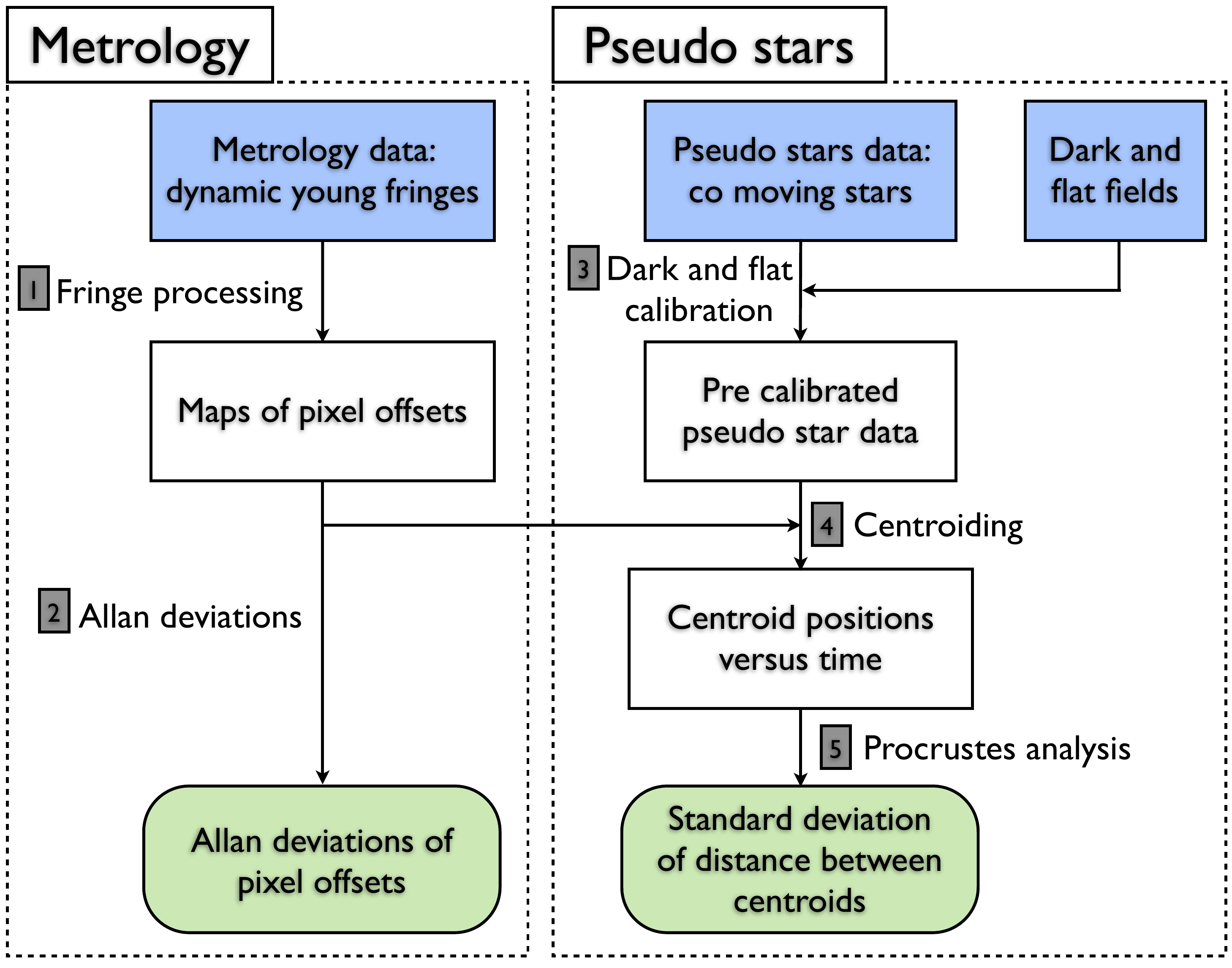}
\caption{\label{schematic_neat_demo_data_processing}\textbf{Overview of data reduction process for the NEAT testbed.} Steps 1 to 5 of the process are described in the next sections.}
\end{center}
\end{figure}

The two following sections focus on the metrology and the pseudo stellar sources. Each one consists in a presentation of the system, the data reduction method and the results obtained on simulated and actual data. Figure \ref{schematic_neat_demo_data_processing} is a diagram summarizing the different types of calibrations involed, it shows how the metrology and pseudo star signal processing are linked together.

Before going into the subject, a word about the units we will use to express the performances of the testbed: we mainly express distances and standard deviations (SD) in units of pixels, sometimes implicitly to simplify the notation. We will also use relative standard deviations (RSD), so beware of confusing the two !

\section{METROLOGY}\label{sec:METROLOGY} 

\subsection{Presentation of the metrology}\label{subsec:Presentation of the metrology}

\begin{figure}[t]
\begin{center}
\includegraphics[width = 15cm]{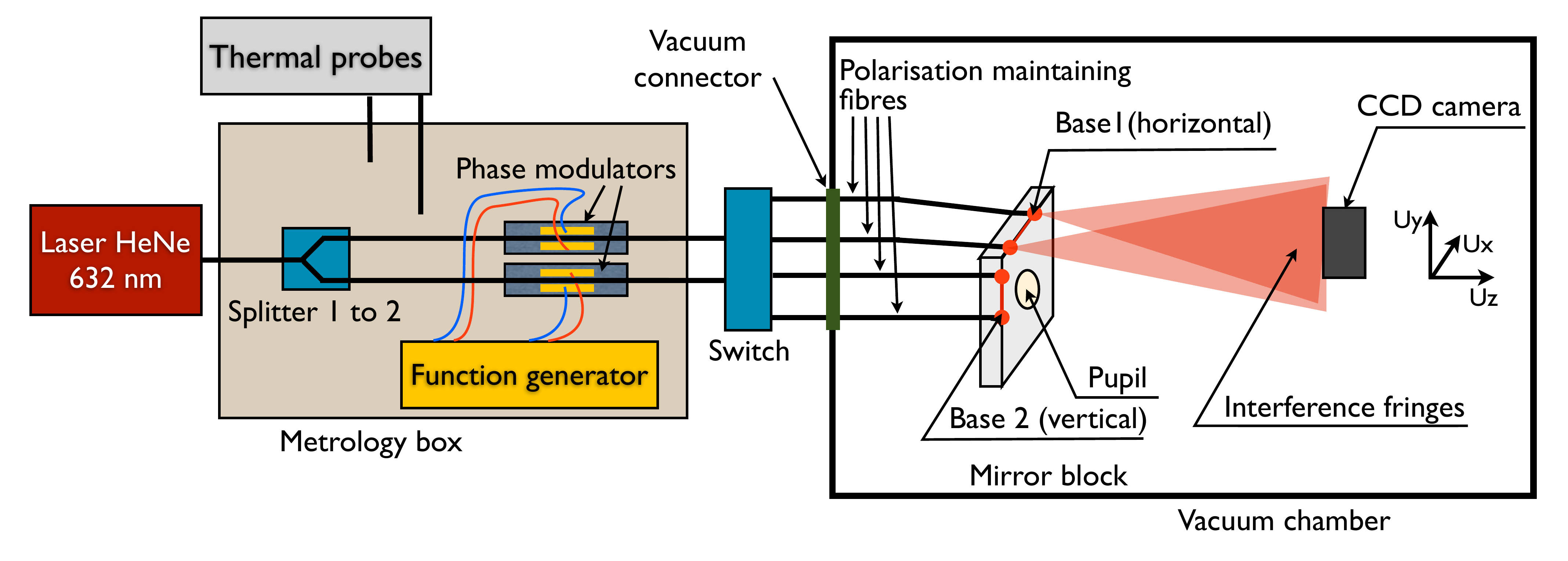}
\caption{\label{fig:metrology_v2}\textbf{Schematic of the metrology.} The axis are indicated on the figure: Z is aligned with the optical axis, X and Y are aligned with the horizontal and vertical directions within the focal plane.}
\end{center}
\end{figure}

The metrology is made of integrated components from the laser to the bases (see Fig.~\ref{fig:metrology_v2}). The source for the metrology is a HeNe laser. Its light is split into two fibers, which are fed into two lithium niobate phase modulators to apply a periodic phase shift between the lanes. This configuration ensures that the phase modulation is applied between the two point sources constituting each base. A switch can be used to select any combination of one or zero fibers on the mirror and most combinations of two fibers. During the metrology calibration phase, two vertically and horizontally aligned fibers are selected successively to project respectively horizontal and vertical dynamic Young fringes.

\subsection{Data reduction methods}

The interference pattern created at the CCD by a given metrology baseline $B$ of coordinates $(B_x,B_y)$ is:
\begin{equation}\label{interference_pattern}
I(x,y,t) \propto 2I_0\left[1+V\cos\left(\phi_0 + \Delta \phi(t) + \frac{2\pi(x B_x + y B_y)}{\lambda_{\ml{met}} L}\right)\right]
\end{equation}
Where $I_0$ is the average intensity at the focal plane for one fiber, $L$ is the distance between the fibers and the CCD, $\phi_0$ is a static phase difference, $\Delta \phi (t)$ is the modulation applied between the lines, $\lambda_{\ml{met}}$ the wavelength of the laser source. Although the exact shape of the fringes is hyperbolic, at the first order the fringes are straight and aligned with the direction perpendicular to the metrology baseline. If we also assume that the point sources are of equal intensity and that the intensity created at the focal plane is uniform we have a fringe visibility of $V=1$. In reality, the visibility is affected by the intensity and polarization mismatch between the point sources. Each fiber project a Gaussian beam whose intensity is spatially dependent and the beams are not co-centered. 
\\

Because all pixels see different visibilities and different average intensities, the solution for the cube of metrology data is searched under the following form:
\begin{equation}
I(i,j,t) = B(t)\iota(i,j) + A(t)\alpha(i,j) \sin\left[iK_x(t)+jK_y(t) + \phi(t) + \delta_x(i,j)K_x(t) + \delta_y(i,j)K_y(t)\right]
\end{equation}
Where $i,j$ are integer pixel position indexes and $\delta_x$ and $\delta_x$ are pixel offsets, i.e. the difference between the pixel true locations and an ideal regularly spaced grid. Here we have decoupled time and spatial variations. Also, we have implicitly transformed $t$ into a discrete index representing a frame number. We will keep using this notation for consistency and because it naturally carries the connotation of a dimension associated with time. Also because there are degeneracies between the variables, we add the following constraints: $\sum_{i=1}^{n} \sum_{j=1}^{m}\iota (i,j) = nm$, $\sum_{i=1}^{n} \sum_{j=1}^{m}\alpha (i,j) = nm$, $\sum_{i=1}^{n} \sum_{j=1}^{m}\delta_x (i,j) = 0$, $\sum_{i=1}^{n} \sum_{j=1}^{m}\delta_y (i,j) = 0$, $\sum_{i=1}^{n} \sum_{j=1}^{m} i\delta_y (i,j) = 0$, $\sum_{i=1}^{n} \sum_{j=1}^{m} j\delta_y (i,j) = 0$, in other words, we force the mean values of $\iota$ and $\alpha$ to 1, the mean pixel offset to zero and the offset gradient to zero. The table \ref{metrology_analysis_variable} summarize what is the meaning of each variable and which kind of noise it will absorb. 

\begin{table}[h]
\caption{\label{metrology_analysis_variable} Table of metrology variables for data analysis.}
\begin{center}
    \begin{tabular}{ | l | l | l |}
    \hline
Notation & Name & Absorbed noises \\ \hline \hline
$B(t)$&	Average intensity & Laser flux, common dark/offset fluctuations\\ \hline
$A(t)$& Amplitude & Laser flux and polarization fluctuations\\ \hline
$K_x(t)$& Metrology wavevector ($x$ component) & Laser freq. fluctuation, thermal expansion \\ \hline
$K_y(t)$& Metrology wavevector ($y$ component) & Laser freq. fluctuation, thermal expansion \\ \hline
$\phi(t)$& Differential phase & phase jitter (thermal and mechanical) \\ \hline
$\iota(i,j)$& Pixel relative intensity & Pixel QE variations\\ \hline
$\alpha(i,j)$& Pixel relative amplitude & Pixel QE variations, visibility space dependence\\ \hline
$\delta_x(i,j)$ & Pixel offsets ($x$ component) & -  \\ \hline
$\delta_y(i,j)$ & Pixel offsets ($y$ component) & -  \\ \hline
    \end{tabular}
\end{center}
\end{table}

\begin{figure}[t]
\begin{center}
\includegraphics[height = 90mm]{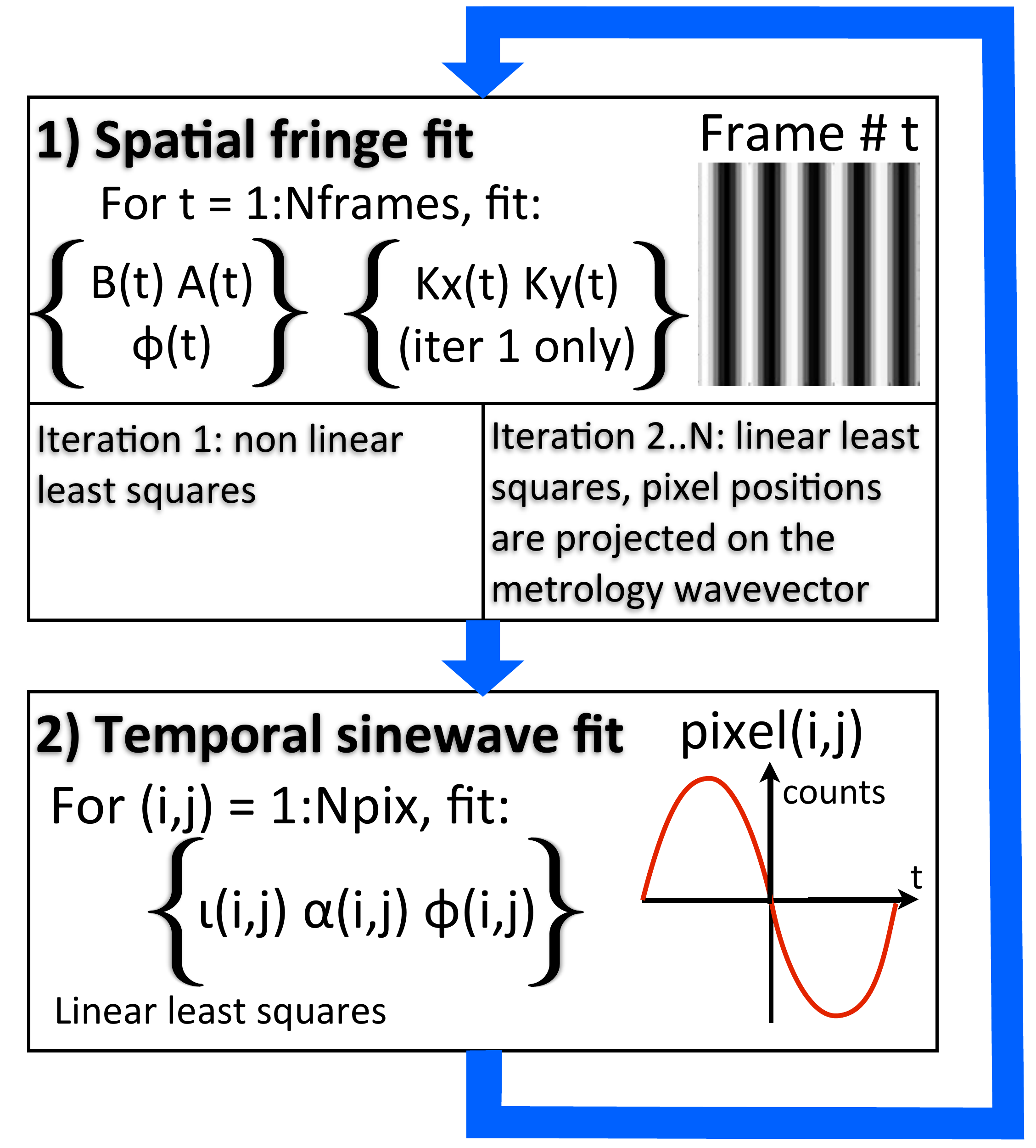}
\caption{\label{metrologyFringeFit}\textbf{Iterative process used to fit the metrology fringes (step 1 on Fig. \ref{schematic_neat_demo_data_processing}).} The difference between the measured phase of a pixel ($\phi(i,j)$) and the phase expected (global fringe phase) is the phase offset caused by the pixel offset projected in the direction of the wavevector.}
\end{center}
\end{figure}

A set of metrology data cube has a size of about 80$\times$80$\times$number of frames, a large cube can contain as many as 200000 frames. The problem is non linear as the fringe spacing is a free parameter. The total number of fitted parameters (80$\times$80+5$\times$Number of frames) is not practical for a straightforward least square minimization of the whole cube, that is why we use an iterative process. First a spatial fit is done on each frame to constrain the time dependent variables and then a temporal fit is done on each pixel to constrain the space (or pixel) dependent variables (see Fig. \ref{metrologyFringeFit}).

The spatial fit is a non linear least square minimization for the first iteration, it uses a levenberg marquart optimisation procedure. The temporal sinewave fit is always a linear one. The method for the temporal fit is very similar to the standard linearizion technique used on a sinewave whose period is known. The difference is that we instead know the phase of the 2D carrier sinewave (not including pixel offsets) at each pixel and for each frame. After normalization to average frame intensity and average frame amplitude, the signal is a pure sinewave with a constant offset whose phase carries information on the pixel location projected along the modulation direction:
\begin{equation}
I(i,j,t) = \iota(i,j) + \alpha(i,j) \sin\left[\phi(t) + \phi(i,j)\right]   = a_{i,j} \sin(\phi(t)) + b_{i,j} \cos(\phi(t)) + c_{i,j}
\end{equation}
Just like in the usual case, a least square minimization of the sum:\\ 
$S_{i,j} = \sum_{t=0}^{N-1} \left[I(i,j,t)- a_{i,j}\sin(\phi(t)) - b_{i,j}\cos(\phi(t)) - c_{i,j}\right]^2$ yields the values for $a$, $b$ and $c$ from which $\alpha(i,j)$, $\phi(i,j)$ and $\iota(i,j)$ are derived for each pixel. $\phi(i,j) = iK_x(t)+jK_y(t) + \delta_x(i,j)K_x(t) + \delta_y(i,j)K_y(t)$ contains the information on the pixel location. Also knowing the metrology wavevector $(K_x(t), K_y(t))$, we derive the projected offset: $\delta_x(i,j)K_x + \delta_y(i,j)K_y$ (for this final step $K$ is assumed constant).

\begin{figure}[t]
\centering
\includegraphics[width = 80mm]{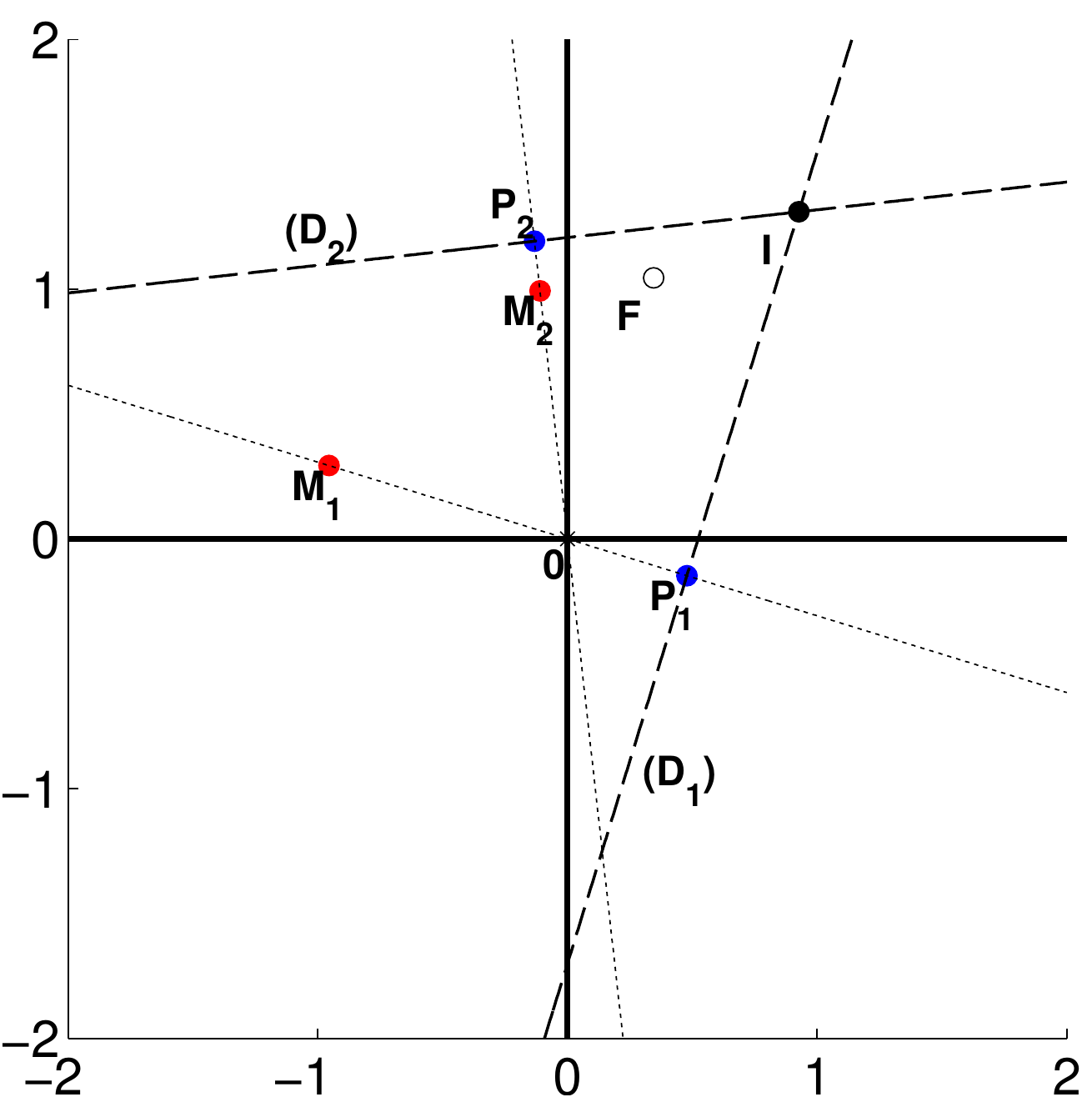}
\caption{\textbf{Illustration of the deprojection problem.} $\protect\overrightarrow{OM_1}$ and $\protect\overrightarrow{OM_2}$ are the wave vectors normalized to a length of one pixel. The points $P_1$ and $P_2$ are the measured offsets, i.e. the projections of the true pixel offset (point $I$), unto the lines generated by the wave vectors. The axis units are in pixels. Simply summing the projected offsets gives a wrong answer (point $F$).}
\label{deprojection_schematic}
\end{figure}

At this point, it is very important to understand that everything we have described about the metrology reduction process applies to a single set of data with a quasi constant $(K_x,K_y)$ metrology wavevector (i.e. the modulation direction of the fringes). The values of $\delta_x$ and $\delta_y$ are internal to the iteration process are not the true pixel offsets, the vector $\delta_x \overrightarrow{U_x} + \delta_y \overrightarrow{U_y}$ is equal to the pixel offset \textbf{projected} unto the metrology wavevector: $\delta_r\overrightarrow{K_r}$. To solve the degeneracy, we repeat the iterative analysis presented above on two sets of metrology fringes (with noncolinear wavectors): two map of projected pixel offsets are obtained. The wavelength vectors of each data set are not strictly perpendicular but fairly close in practice. From this two maps, true $x$ and $y$ offsets (i.e. coordinates in a standard orthonormal basis) are derived by solving for each pixel the intersection between the projected offset coordinates ($\delta_{r,1} \delta_{r,2}$). Figure \ref{deprojection_schematic} illustrates this ‘‘deprojection" problem, the solution (not detailed in this paper) is found by straightforward application of Euclidean geometry in a plane.

\begin{figure}[t]
\centering
\includegraphics[width = 80mm]{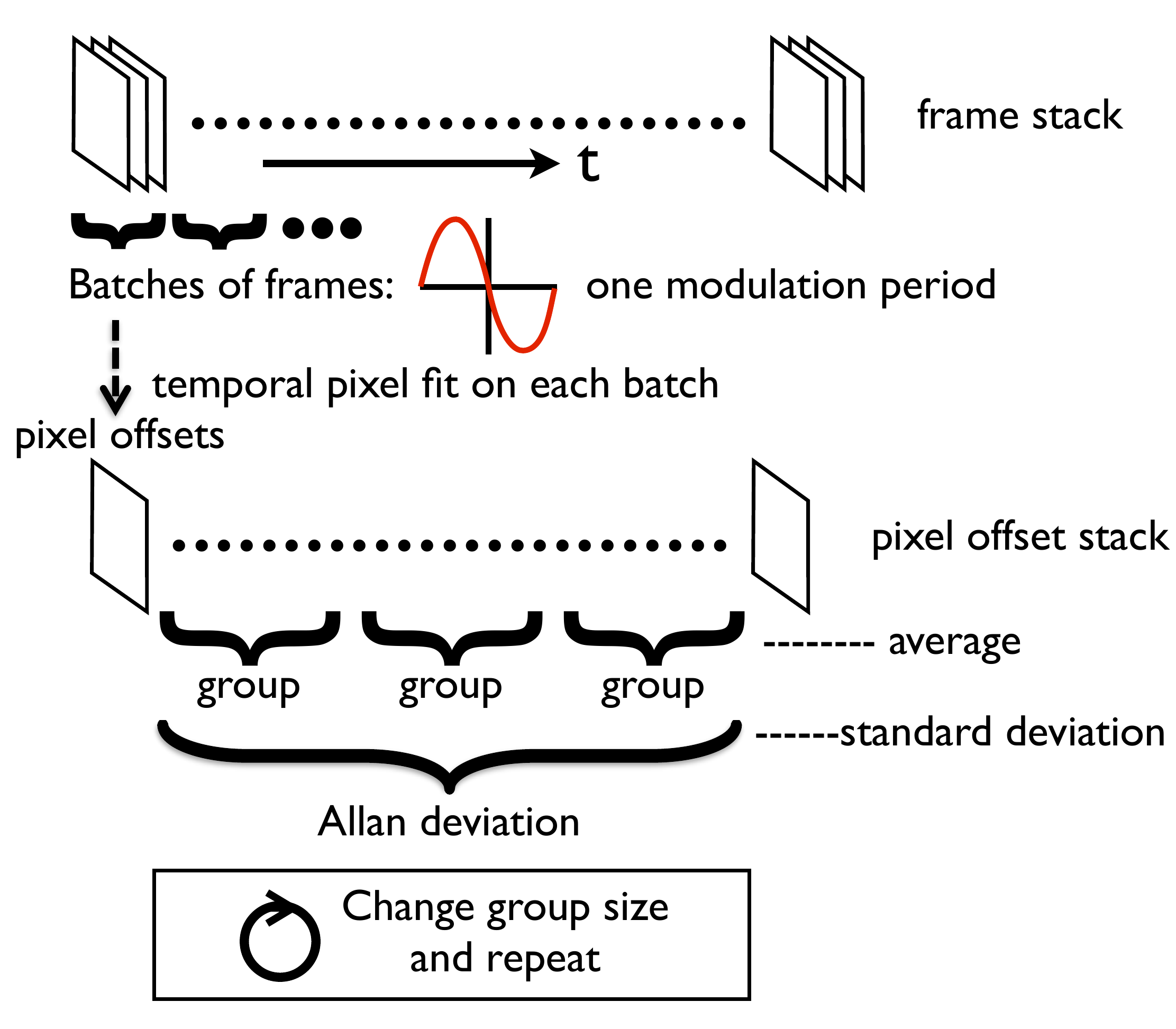}
\caption{\textbf{Diagram of the process used to calculate the Allan deviations (step 2 on Fig. \ref{schematic_neat_demo_data_processing}).}}
\label{allan_deviation_diagram}
\end{figure}

The second important result given by the metrology, besides the true pixel offsets, is the Allan deviation. The latter can be calculated directly on the normalized projected pixel offsets. Our interest here is to estimate the accuracy of the pixel positions we have obtained. This estimation is almost insensitive to the basis in which the coordinates are expressed, as long as the vectors of the basis are close to orthonormal. Figure \ref{allan_deviation_diagram} shows how the Allan deviations are calculated.

The starting point is the dark subtracted frames and the final solution of the spatial fits from the previous iterative process. Using these as starting values, the temporal pixel fits are performed on small parts of the data called batches, instead of the whole data cube. The number of frames in each batch is calculated so that the temporal signal seen by each pixel covers at least one sinewave period: this is needed to have a well constrained fit. One obtains a map of the pixel offset for each batch, this was the first step.

For the second step, the Allan deviations per say are applied on the cube of pixel offsets. The principle is to form groups of pixel offsets maps (in fact group of batches), to calculate the average within each group and then the standard deviation between the groups. The final standard deviation depends on how many batches each group has. 

Each group corresponds to one point on the Allan deviation plot, so the second step is repeated for different group sizes. The maximum group size is when the standard deviation is calculated on only two groups. The point of doing all this is to simulate doing several experiment and looking how the accuracy changes with the experiment duration. By splitting the data into several subsets, only one experiment is needed.

\subsection{Results on simulations}

\begin{figure}[t]
\centering
\includegraphics[height = 109mm]{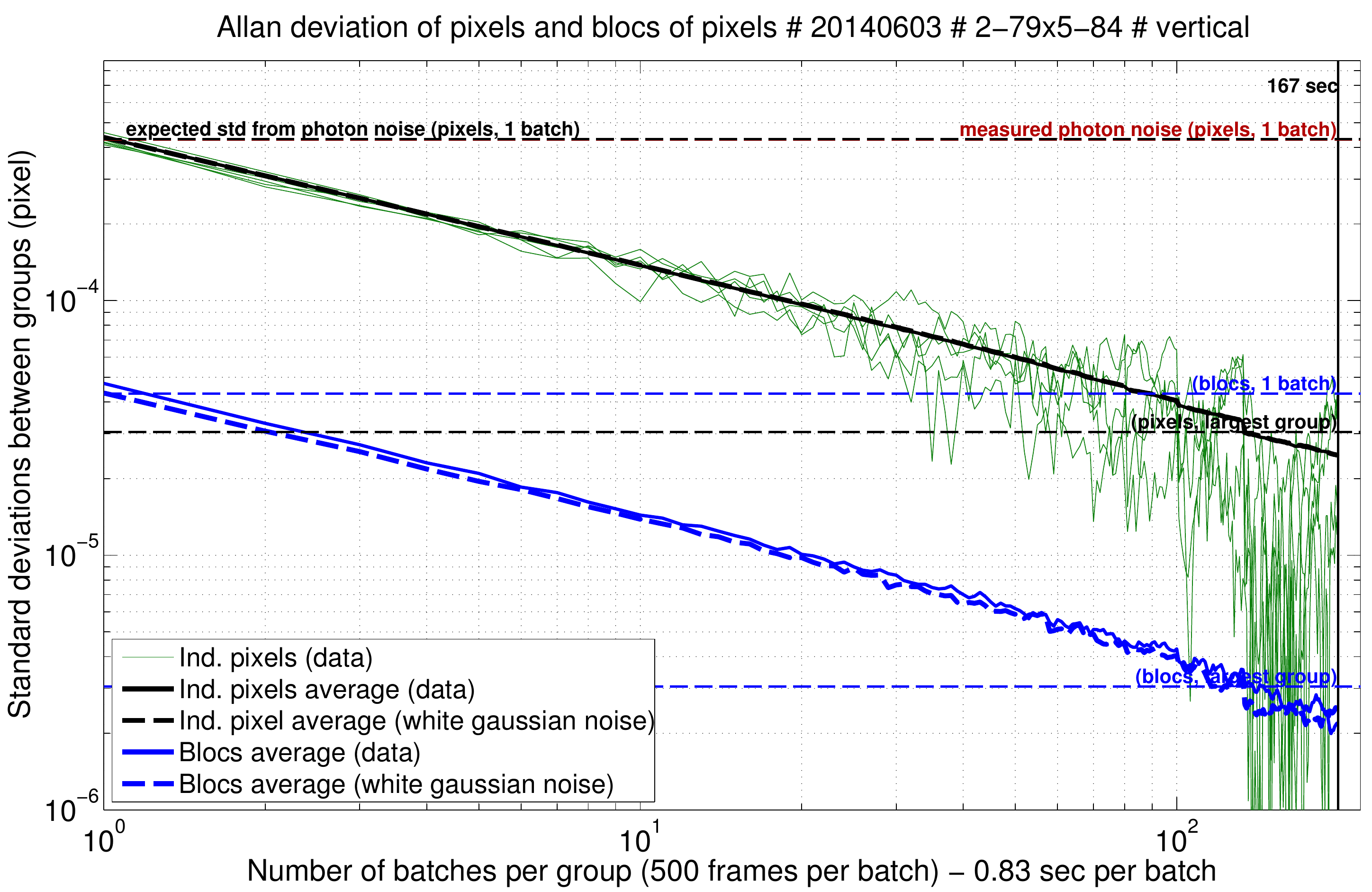}
\caption{\textbf{Allan deviations of simulated data.} The amplitude, visibility and photon noise of simulated fringes are adjusted to values typical of real experiment (B = 10000 counts, A = 6000 counts, 1 count = 10 photons). Additional sources of noise are simulated, such as laser intensity ($1\e{-2}$ RSD), fringe phase ($1\e{-2}$ radian SD) and pixel QE ($1\e{-5}$ RSD). The plot shows deviations for individuals pixels (plain red), their average deviation (plain black) and the average for blocs of 10 by 10 pixels in (plain blue). The dotted black and blue  curves are for a cube of white noise whose standard deviation is matched to the data for groups of 1 batch. Averaging Allan deviations over pixels or blocs is important because they tend to be noisy (plain red) when the final deviation is derived from very few groups. Horizontal dotted blue and black lines are various estimations of the photon limit.}
\label{allan_deviation_simData_example}
\end{figure}

Figure \ref{allan_deviation_simData_example} shows the Allan deviation obtained after analysis of 200,000 simulated fringes. The goal is to validate the pipeline and the photon error budget. This number of frames corresponds to about 3 minutes of data with our setup. We can acquire data up to 1000 Hz, but in practice we work somewhere between 600 and 800 Hz because of the intensity limit set by the metrology fibers output.

For groups of 200 batches (i.e. 100,000 frames), the accuracy reaches $2\e{-6}$ for groups of pixels and $2\e{-5}$ for individual pixels. This level of accuracy is entirely satisfactory (see section \ref{subsec:PSEUDO STELLAR SOURCES} to understand why). It also means that our previous photon error estimation\cite{Crouzier12} ($N_{\ml{ph}} = 1/\sigma^2$ where $\sigma$ is the deviation in pixel units) was roughly correct, within a factor of two (the needed number of batches per group to reach $2\e{-5}$ is 83 according to this formula). Since then we have refined our estimation of the propagation of photon noise unto the pixel location but we will not detail the method here. The expected deviations are indicated by horizontal dashed lines on the plot and they coincide almost perfectly with the measured Allan deviations. The top black line shows what the deviation for groups of one batch should be: as expected, it crosses the left axis of the plot at the same place than the Allan deviation curve (black line). They are actually two lines near the top, almost on top on each other: the dark one is for theoretical photon noise, the red one is for measured photon noise using the first frame of the data cube.

\begin{figure}[!t]
\centering
\subfigure[]{\label{pixelOffsetBias_pixelQE1e-5RS}
\includegraphics[width=58mm]{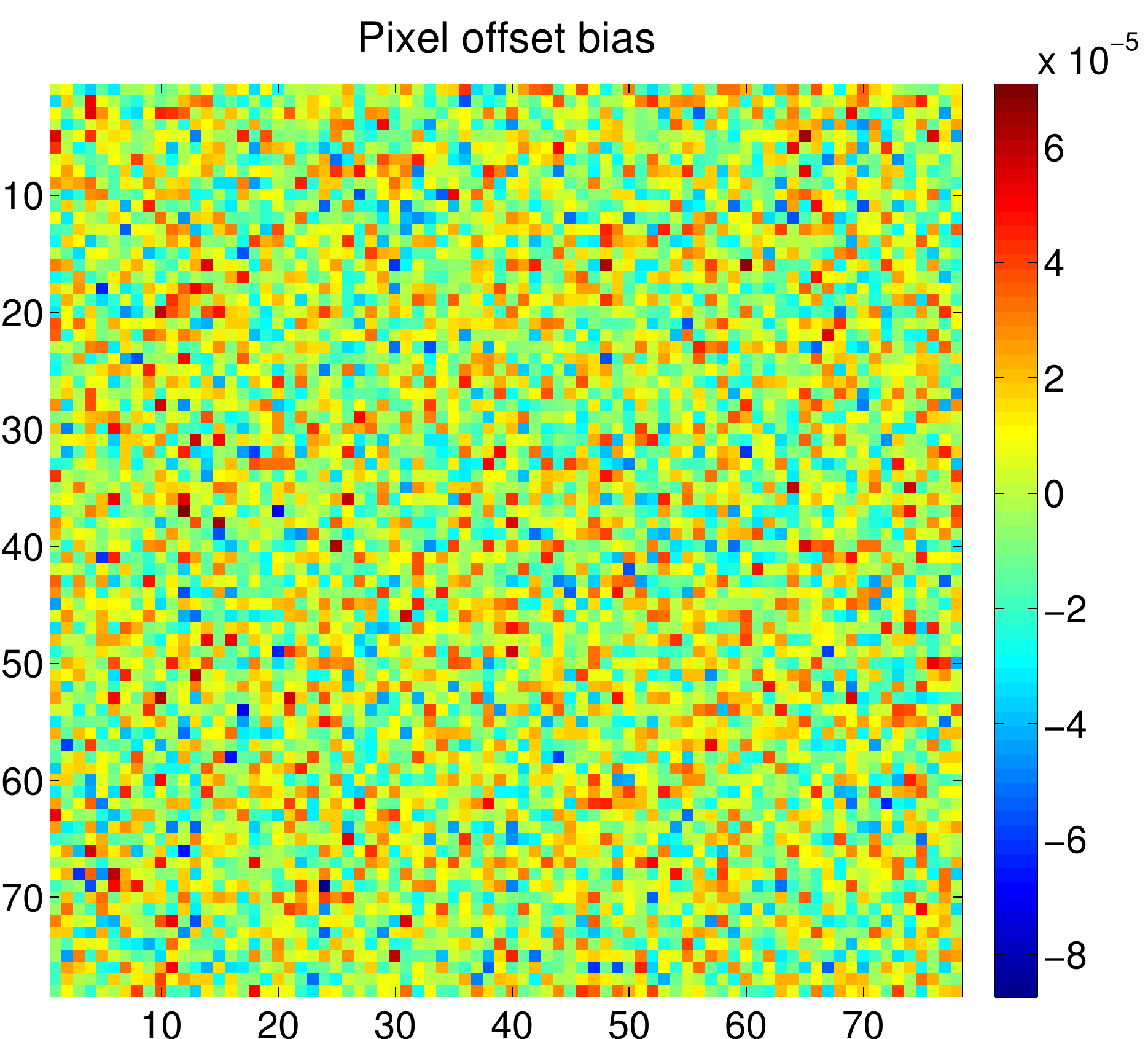}}
\hspace{5pt}
\subfigure[]{\label{pixelOffsetBias_pixelQE1e-2RS}
\includegraphics[width=58mm]{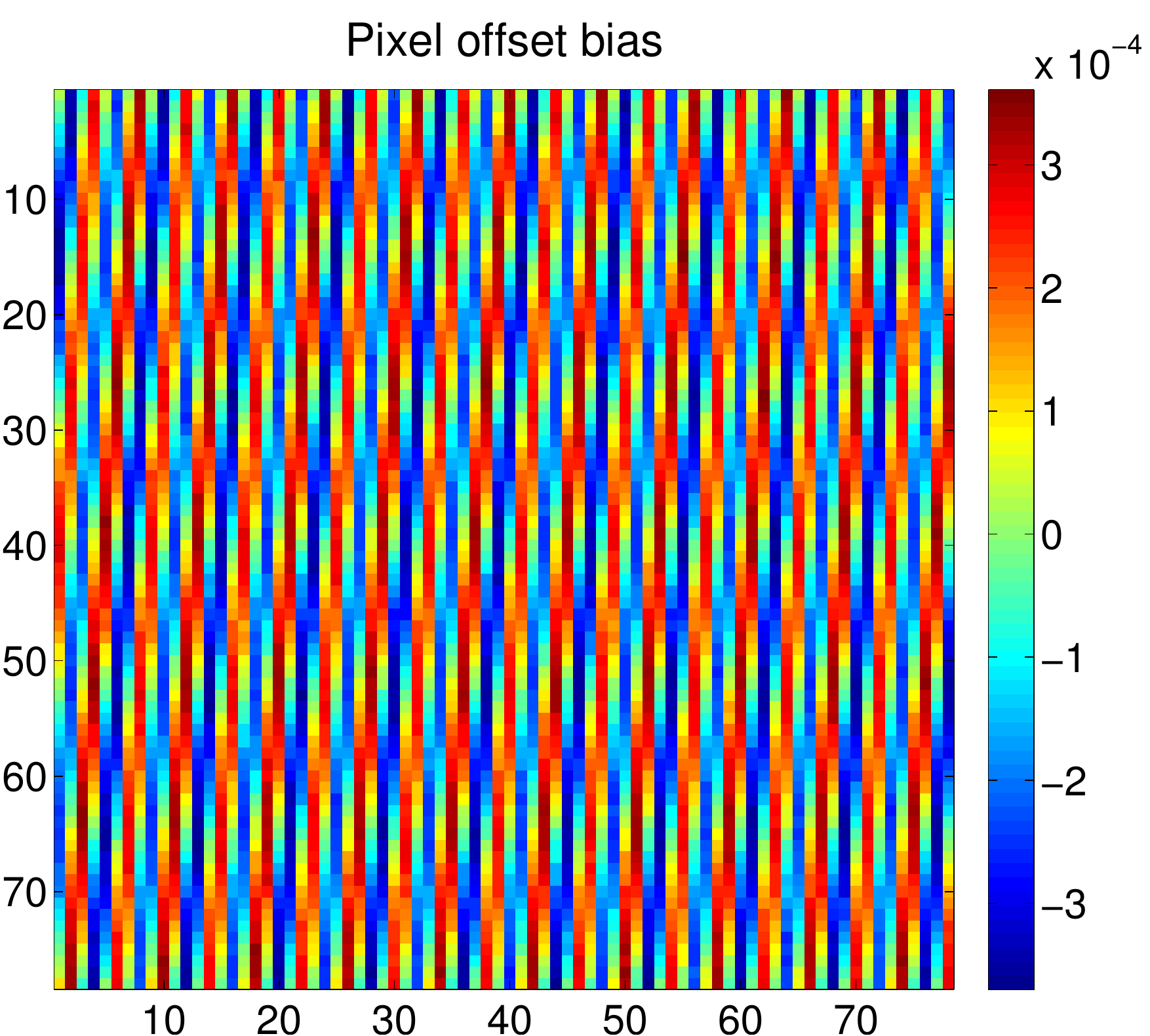}}
\caption{\label{pixelOffsetBias}\textbf{Pixel offsets bias for different pixel QE RSD.} The maps show the difference between the pixel offsets found after reduction and the solution of the simulation, for pixel QE RSD of $1\e{-5}$(a), and $1\e{-2}$ (b). The residuals for (a) are in agreement with the photon noise floor, while a systematic bias of a few $1\e{-4}$ pixels is visible for (b). Both have same photon noise.}
\end{figure}

Figure \ref{pixelOffsetBias} shows maps of the difference between the measured pixel offsets and the true solution (pixel offset simulation input) for different values of pixel QE RSD. This is the ultimate metric to check the accuracy of the result, because biases constant in time are not visible in the Allan deviation. For low pixel QE RSD ($<1\e{-5}$), the standard deviation of $2\e{-5}$ seen in Fig. \ref{pixelOffsetBias} is coherent with the value given by Allan deviation (i.e. the photon noise floor). However, tests with higher values of pixel QE RSD show residual spatial bias above photon noise floor, while the Allan deviation is unaffected. This is an eventual point to improve, but a good pre-flat calibration of the actual metrology data can in principle take care of this problem.

In conclusion, although a small bias could persist if the metrology data is not properly flat calibrated (ideally to an accuracy of $1\e{-5}$), the Allan deviations are not affected and are a good way to control for the presence of correlated noise. The comparison between simulations and actual data will give us clues about what kind of noises are present in the data.

\subsection{Results on data}

Various experiments have been done to perform metrology calibrations on a wide range of operating conditions: in air (CCD temperature stabilized at 17$\dg$C), in vacuum (P $<$ 0.5 mbar, CCD temperature stabilized at 17$\dg$C or down to -10$\dg$C) and with a different setup for the control of stray light. 

\begin{figure}[t]
\centering
\includegraphics[width = 160mm]{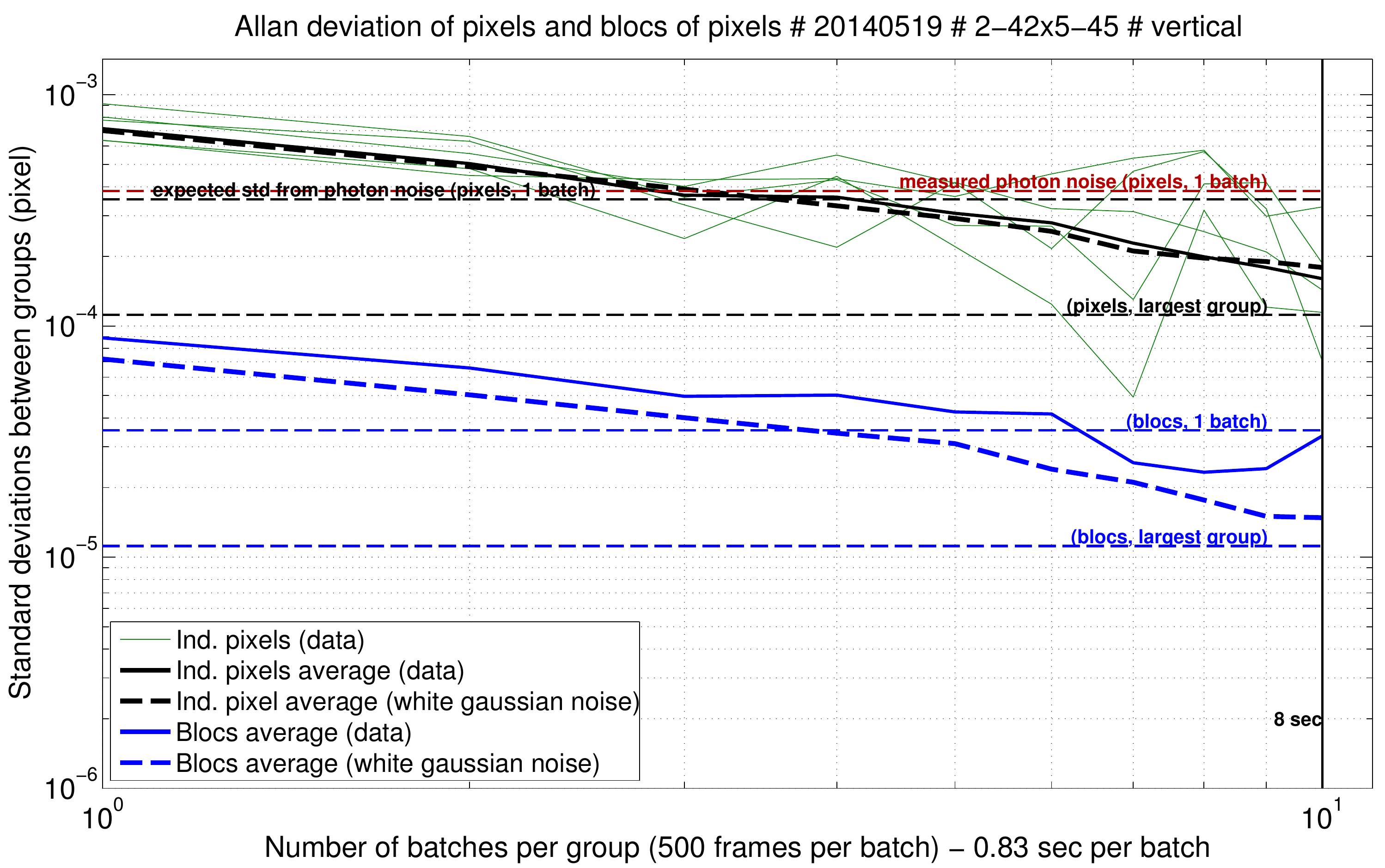}
\caption{\textbf{Allan deviations (CCD data, in air, 17$\dg$C, 10000 frames, using the new baffle).} Only one quadrant was used in this analysis to avoid having pixels from the ‘‘bad quadrant" in the data. They are noticeable differences with the simulated data. Firstly the level of noise is higher than the expected photon noise limit by a factor 2.5. Secondly, the gain of spatial averaging (blocs of pixels) is slightly less the expected factor of 10. Thirdly, tests on larger data sets (up to 50000 frames) show degraded accuracy: the two symptoms mentioned above worsen (worse spatial averaging, larger gap with theoretical photon noise) and the standard deviation does not fall as quickly. All this indicates the presence of spatial and temporal red noise.}
\label{allan_oneQuadTubeBaffle_10000frames}
\end{figure}

The results we present here are only Allan deviations. We will link the metrology and the centroiding accuracy in the next section. The results have so far not been very sensitive to changes in operating pressure or temperature. The best performances were reached with a set of data taken in air at 17$\dg$C, for which we have selected the upper left quadrant only. The CCD is divided into 4 quadrants of 40 by 40 pixels, each one having separate readout electronics. One of the quadrants showed a time dependent systematic behavior, with a standard deviation on the pixel offsets up to 10 times higher than the rest of the CCD. The cause of this behavior is still unknown and under investigation, in the meantime we have simply restricted our analysis to a part of the CCD unaffected by the problem. The resulting Allan deviation is shown by Fig. \ref{allan_oneQuadTubeBaffle_10000frames}.

We have also have looked at the impact of stray light on Allan deviations and found interesting results. A test without the baffle (all other things constant: one quadrant, 10000 frames) showed that the stray light has a strong impact: Allan deviation of $3\e{-4}$ for pixels and $2\e{-4}$ only for groups of pixels. Tests with a baffle made of paper sheets have shown similar result than with no baffle.

We will now briefly comment on the evolution compared to the results presented in the last SPIE paper. The previous Allan deviations were projected on the X and Y axis whereas the current results are always projected along the modulation direction, where the sensitivity is maximum. The fringes are slightly inclined: there is a major projection component and a minor one that is about 10 times smaller. Now that we have figured out the deprojection process, we understand that previously we have naively and incorrectly selected the (best looking) minor axis. For a meaningful comparison, the relevant pixel offsets measurements have to be scaled about 10 times higher, and so does the Allan deviation, so the previous result (summer 2013) was in fact closer to $1\e{-3}$ (for blocs of 11 by 11 pixels) and not $1\e{-4}$ as claimed. By comparison, the best current result is $3\e{-5}$ for groups of 10 by 10 pixels. Although very significant, the difference is hard to interpret, it can be caused by a lot of changes in operating conditions and by a better processing. Numerous improvements have been made on data processing along the way, such as validation on simulated data.

\begin{figure}[!t]
\centering
\includegraphics[height = 50mm]{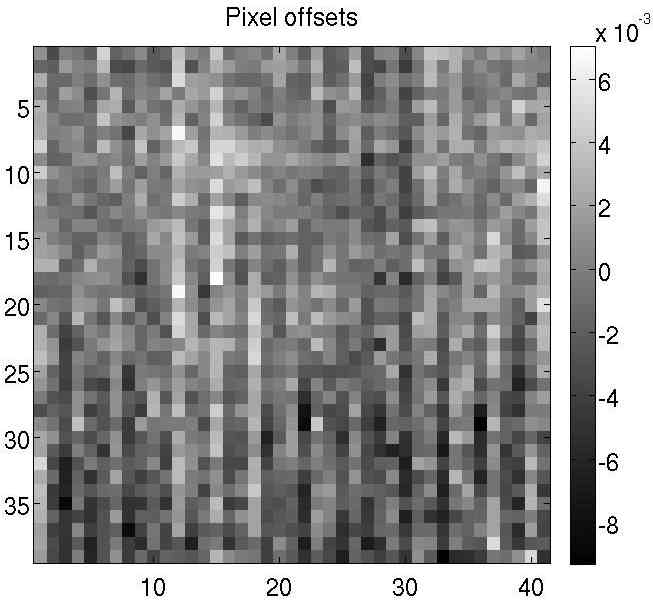}
\caption{\textbf{Pixel offsets for quadrant 1.} The magnitude of the offsets is between $+7$ and $-9\e{-3}$ pixels, corresponding to a SD of about $2.6\e{-3}$ pixels. We have used vertical fringes: the offset measured is the projection of the true offset in a direction close to horizontal.}
\label{pixelOffsetsQ1}
\end{figure}

Here as a conclusion Fig. \ref{pixelOffsetsQ1} shows the pixel offset map that corresponds to the best Allan deviation analysis (one quadrant only).

\section{PSEUDO STELLAR SOURCES}\label{subsec:PSEUDO STELLAR SOURCES}

\subsection{Presentation of the pseudo stellar sources}\label{subsec:Presentation of the pseudo stellar sources}

\begin{figure}[t]
\begin{center}
\includegraphics[height = 50mm]{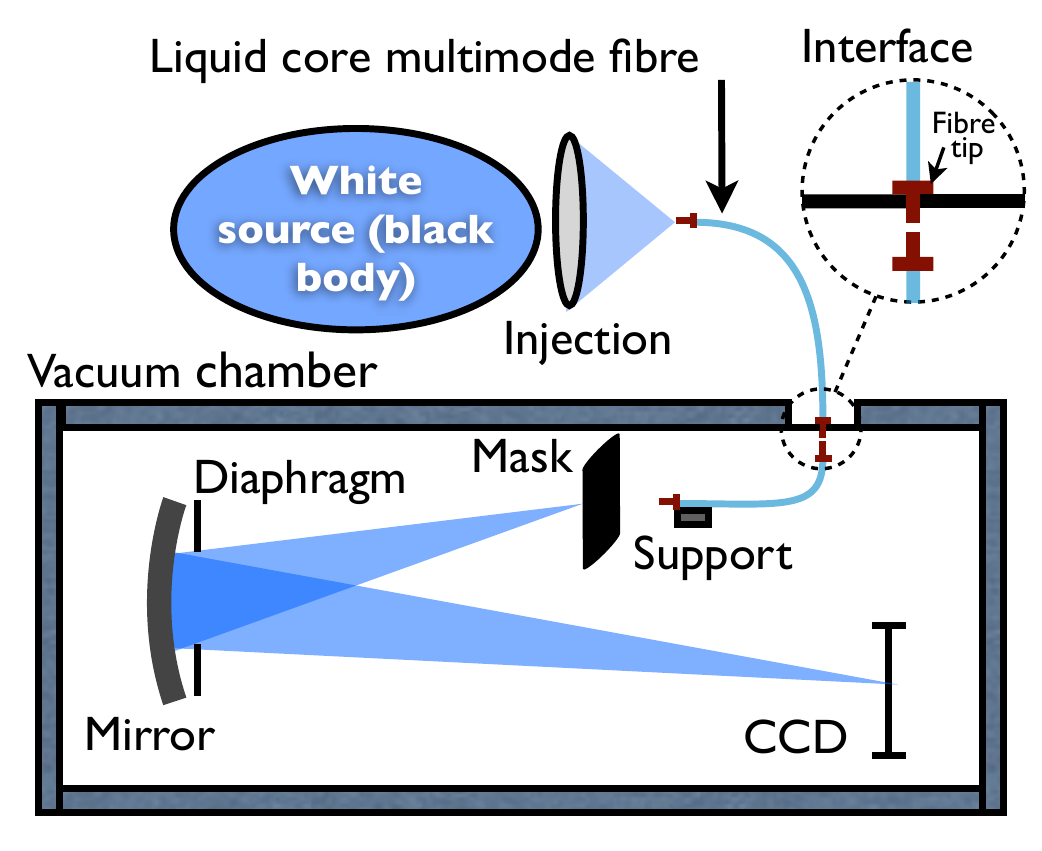}
\caption{\label{fig:scheme_pseudostellar_sources}\textbf{Schematic of the pseudo stellar sources.} We used a magnification factor of 2 and an off axis angle of 2 degrees. This configuration allows the installation of the pseudo stellar sources and the camera without any beam obstruction with some margins to accommodate the support elements. Additionally, with an aperture as small as 5 mm, a spherical surface is sufficient to obtain optical aberrations that produce a spot diagram smaller than the diffraction pattern in the whole field of view.}
\end{center}
\end{figure}

The pseudo stellar sources system function is to project 5 stars unto the CCD. The 4 outer stars represent reference stars, the central star is the target. This star configuration allows us to perform a precise differential measurement: XY position offset and scale changes can be measured between the stars. In order to use a reasonable approximation of the spectrum of a real star, we use a black body source. The schematic of the system is presented by Fig. \ref{fig:scheme_pseudostellar_sources}.


\subsection{Data reduction methods}

\begin{figure}[t]
\begin{center}
\includegraphics[height = 80mm]{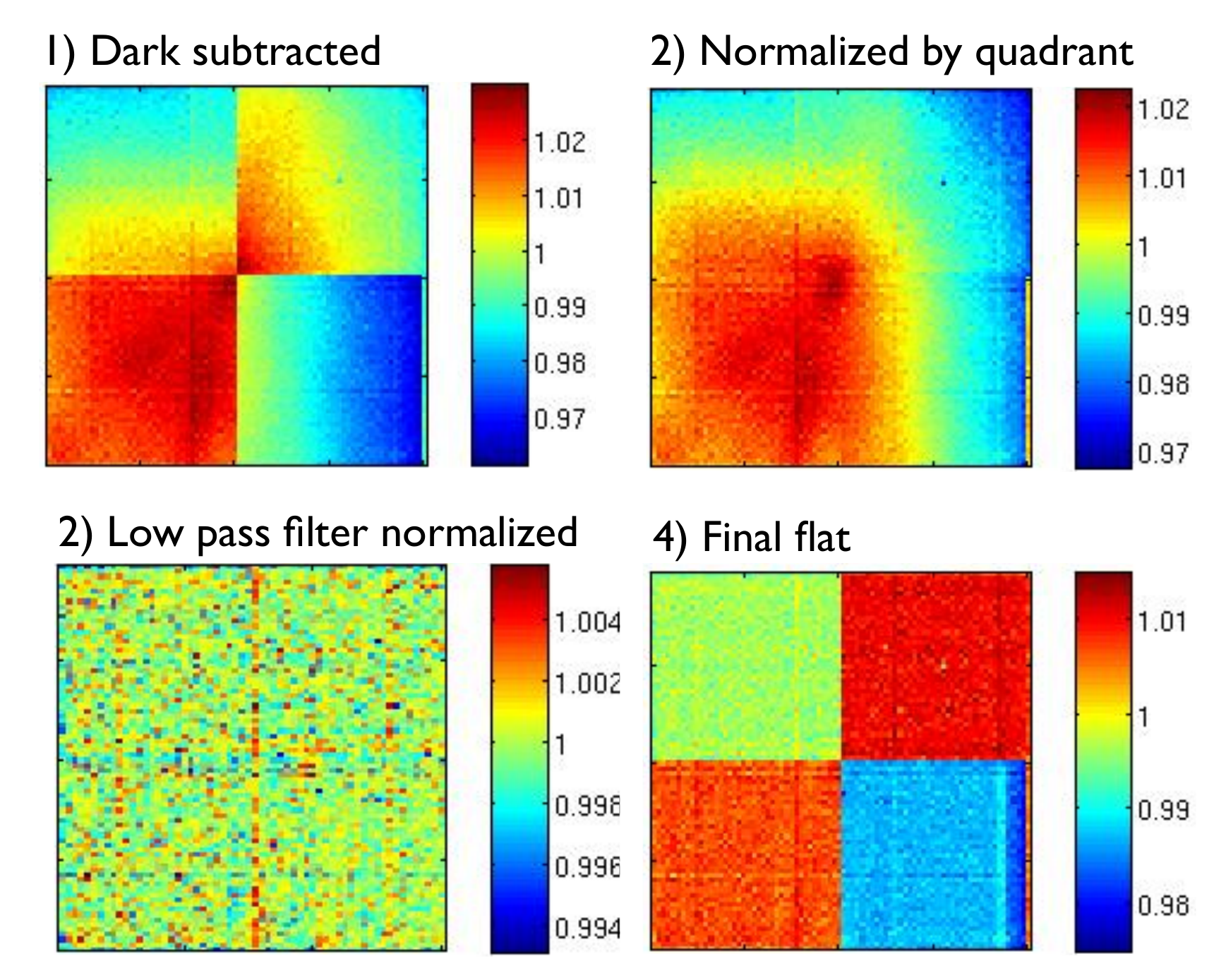}
\caption{\label{flatProcessing}\textbf{Flat field processing.} After dark subtraction (1), the relative gain between the quadrants are computed assuming continuity of the intensity between the quadrants. Image (2) is then divided by its low pass filter component to obtain (3). At last, the quadrant relative gains are added back into the data to obtain (4) which is the flat that will be used for the centroids or the metrology.}
\end{center}
\end{figure}

The first step of the data reduction is the standard dark subtraction and flat calibration. The application to the data is straighforward: the reduced data cube is simply: $I' = (I - \ml{dark})/\ml{flat}$, but the delicate part here is obtaining a high quality flat field in the first place. The vacuum chamber doesn't allow for an integrating sphere, it would bloc the light from pseudo stars and the metrology: we have to use the metrology fibers. The method consist in turning-on only one metrology fiber, which will produce a fairly flat intensity pattern on the CCD. The intensity profile produced is a Gaussian beam whose waist of about 10cm is much larger than the CCD field (2mm). So what is seen by the detector is an intensity gradient with an eventual slight curvature, which has to be ‘‘detrended". The method used to obtain the flat field is illustrated by Fig. \ref{flatProcessing}.

An important point is that the flat must be done in incoherent light, unlike the metrology fringes. Coherent light produces relative intensity variations $\propto \sqrt{I_1/I_0}$ instead of $\propto I_1/I_0$ because of interferences ($I_0$ is the intensity of the fiber, $I_1$ is the intensity of a parasite reflection). The interference pattern can be quite complex and have spatial features unresolved by the pixels if the angular separation is too large, so it can not be properly detrended. This is the case for example for reflections on the edges of the stop apertures of the baffle.

The second step is the centroiding. So far, we have worked with with two different methods. The first one is a simple Gaussian fit with 7 parameters (background level, intensity, position X, position Y, width X, width Y, angle). The second one is an autocorrelation technique. The principle is to measure a displacement between two images by fitting a phase ramp between the images in the Fourier domain. When done on Nyquist sampled data, it is equivalent to a perfect interpolation. Another advantage of this method is that we are essentially using the data itself to reconstruct the PSF and we do not need a model of the PSF in the first place, thus avoiding potential errors caused by a model/data PSF mismatch. But this method has a drawback: it only gives relative motion. In order to know the distance from one centroid to another, we have to use an autocorrelation between two distinct centroids. However, because we have an optical configuration with only one optical surface and no obscuration of the field of view, the PSF is expected to be nearly invariant. The errors of this process should be mainly caused by the pixels, which can be calibrated by the metrology. 

The third and last step is the SD analysis of the centroid positions, with or without Procrustes analysis. At this point we have a measure of the coordinates of each centroid versus the time, for one or more positions on the CCD. We perform several types of analysis. For the first type (lets call it quasistatic), we monitor the centroids locations versus time at a fixed position: we do not use the motors of the translation stage. We can then calculate the SD of the distances between the outer centroids and the central one. This measure is sensible to some environmental factors such as mechanical stability, air turbulence etc... but if the centroid positions are stable enough, the pixel errors are nearly constant and therefore do not affect the measures significantly. The centroids motion are caused by mechanical vibrations and are typically less than one 1\% of a pixel. A complete Procrustes analysis is actually not needed to obtain this particular result, only the translation is compensated. 

For the second type of analysis (lets call it dynamic), we move the centroids at different positions on the CCD with the translation stage. The amplitude of the motion between each position can be controlled and can range between 1\% of a pixel to several pixels. We also calculate the SD of the distances between the outer centroids and the central one. But this is not enough: because the translation stage induces large tip-tilt errors we have to take into account vertical and horizontal scale changes. In order to calibrate the scales, we use the centroid themselves, by doing a full Procrustes analysis. The principle is to find the geometric transformation that result in the closest overlap of the measured centroid positions. The residuals between the overlaps indirectly yields the final accuracy. The number of parameters needed to define the transformation is 5 (translation X and Y, scaling X and Y, rotation), which is less than the number of data points (2 axes $\times$ 5 centroids for each position). This is what makes the analysis possible.

\subsection{Results on simulations}

In simulations, we have until now approximated the PSF with a Gaussian function whose width is roughly equivalent to an Airy spot at the average wavelength for our experiment (about 600 nm). The reason for using a Gaussian instead of a more complex PSF shape, such as a polychromatic sum of Airy functions or a PSF derived from the data is simplicity and efficiency.

The result from the simulations are used in two different ways. The first goal is to validate the reduction process itself, as capable of reaching $1\e{-6}$ pixel in ideal conditions. We have validated the accuracy in ideal conditions for both centroiding techniques (Gaussian and autocorrelation). Of course validation of Gaussian fitting on Gaussian data does not tell much of the performances on real data, it is nothing more than an advanced debugging method. As it turns out, the accuracy of a Gaussian fit on a real centroid is about a few mili-pixels, but it is still very useful. We will actually use Gaussian fits on real time data to monitor the position of the translation stage and reposition the CCD at an accuracy of 1\% of the pixel for future tests.

The second goal is to explore what are the impact of different types of noise by injecting them one at a time into the simulated data. Because we use Gaussian PSFs, in simulations we can directly compare the input centroid locations with the fit results, thus relying on absolute positions (as opposed to distances) and we can also skip the Procrustes analysis. We have used the model to estimate the relations between the uncertainties on various parameters (pixel sensivities, pixel offsets, ghost centroids) and the centroiding accuracy, the results are summarized in table \ref{result_centroid_model}.

\begin{table}[t]
\caption{\label{result_centroid_model}Results from pseudo stars model. The error on pseudo stars measured locations is always in pixel units.}
\begin{center}
\begin{tabular}{ | l | l | l |}
\hline
Error type 		 							& Error normalization / definition				& Error on centroid   \\ \hline\hline
Pixel sensitivity: $\sigma_{\ml{QE}}$ & average pixel sensitivity = 1 & 0.5$\sigma_{\ml{QE}}$ \\ \hline
Photon noise : $\sigma_{\ml{ph}}$	& \specialcell{relative photon noise\\calculated for the pixel\\with the highest value} & 0.75$\sigma_{\ml{ph}}$\\ \hline
Pixel offset : $\sigma_{\ml{offset}}$ & \specialcell{offset expressed\\in pixel units} & $0.15\sigma_{\ml{offset}}$ \\ \hline
Pixel read noise: $\sigma_{\ml{read}}$ & \specialcell{relative read noise\\calculated for the pixel\\with the highest value} & 1.1$\sigma_{\ml{read}}$ \\ \hline
Ghost centroids: $\sigma_{\ml{ghost}}$ & \specialcell{ghost intensity relative to the main\\centroid, ghost located randomly\\within 2 pixels of the star center} & 1.2$\sigma_{\ml{ghost}}$\\ \hline
\end{tabular}
\end{center}
\end{table}

We can draw interesting conclusions from these relations. In order to reach an error of $5\e{-6}$ on the centroid, one only needs to calibrate the pixels positions to about $3\e{-5}$ and the accuracy of the flat field has to be better than $1\e{-5}$. On the other hand, the experience is very sensitive to the presence of a ghost centroid near the real one: its intensity must be lower than $4\e{-6}$ times the main centroid intensity.

\subsection{Results on data}

On real data, we perform both types of analysis for each setup (quasistatic and dynamic). Each setup produces a data set which consist of several data cubes (one per position of the translation stage). In analysis runs we typically use between 4 and 16 different positions, depending on where we allow the centroid to be placed. Because the CCD is only 80 by 80 pixels with 4 quadrants, we have fewer positions available if we want to avoid putting the central centroid at the limit between two quadrants or moving from one quadrant to another. The quasistatic analysis only need one data cube (one position) to work: having several of them, we simply take the average.

%

Table \ref{result_centroid_data} shows the results for different setups. In each setup there are 3 different types of results: the first one, ‘‘relative accuracy" is a dynamic analysis, without procrustes analysis, the second one, ‘‘relative accuracy (with PS)" is a dynamic analysis, with Procrustes superimposition and the third one, ‘‘relative precision" is the quasistatic analysis. We are dealing with \emph{relative} measurements: centroids positions are always measured relative to one another to absorb common motions like mechanical vibrations. The expected photon noise on the centroids is $1.0\e{-5}$ in all 5 cases because it depends on the batch size and on centroid intensity which both barely change. Other important parameters that are common to all the cases presented are: the PSF fitting method (autocorrelation), the size of the window used for the fit (9 by 9 pixels) and the fact that there is no correction for the pixel offsets. In order words, we have not used the metrology measurements to improve the centroiding accuracy.

\begin{table}[h!]
\caption{\label{result_centroid_data}Results from pseudo stars data.}
\begin{center}
\begin{tabular}{ | l | l |}
\hline
Data/analysis configuration	& Results \\ \hline\hline
\specialcell{2014/01/23 \# 130000 frames \# Batch size: 28000\\Flat calibration: none\\Optical table: OFF} 
& \specialcell{relative accuracy: $1.05\e{-2}$\\relative accuracy (with PS): $1.66\pm0.23\e{-3}$\\ relative precision: $9.73\pm3.3\e{-5}$} \\ \hline
\specialcell{2014/01/23 \# 130000 frames \# Batch size: 28000\\Flat calibration: HeNe laser @633 nm\\Optical table: OFF} 
& \specialcell{relative accuracy: $1.03\e{-2}$\\relative accuracy (with PS): $2.02\pm0.28\e{-3}$\\ relative precision: $9.96\pm3.4\e{-5}$} \\ \hline
\specialcell{2014/01/31 \# 80000 frames \# Batch size: 26000\\Flat calibration: none\\Optical table: ON} 
& \specialcell{relative accuracy: $6.75\e{-3}$\\relative accuracy (with PS): $1.45\pm0.2\e{-3}$\\ relative precision: $1.49\pm0.7\e{-5}$} \\ \hline
\specialcell{2014/01/31 \# 80000 frames \# Batch size: 26000\\Flat calibration: white light\\Optical table: ON} 
& \specialcell{relative accuracy: $6.84\e{-3}$\\relative accuracy (with PS): $1.25\pm0.17\e{-3}$\\ relative precision: $1.54\pm0.7\e{-5}$} \\ \hline
\specialcell{2014/01/31 \# 80000 frames \# Batch size: 26000\\Flat calibration: SLED @740nm\\Optical table: ON} 
& \specialcell{relative accuracy: $6.85\e{-3}$\\relative accuracy (with PS): $1.19\pm0.16\e{-3}$\\ relative precision: $1.50\pm0.7\e{-5}$} \\ \hline
\end{tabular}
\end{center}
\end{table}

For the first two sets (2014/01/23), the central centroid was placed to 4 positions, each one on a different quadrant. For the other 3 data sets, it was moved to 4 places on the same quadrant, the 4 locations being closer from one another. This explains why the accuracy before Procrustes is better in the three last cases (less tip-tilt). From table \ref{result_centroid_data}, we draw the following conclusions:
\begin{itemize}
\item Turning the optical table on (i.e. pneumatic damping) decrease the amplitude of the vibrations from about 1\% of a pixel to a few thousands of a pixel. This explains why the relative precision is closer to the photon noise limit for the last three cases: the bias caused by the pixels are much smaller.
\item With our setup, plate scale variations cause errors of about $1\e{-2}$ pixel. As the stars are separated by 28 pixels, this corresponds to a plate scale variation of $3\e{-4}$. Procrustes analysis is mandatory to go beyond 0.01 pixel.
\item The flat calibration with the Hene laser (monochromatic light at 633 nm) actually deteriorate the accuracy. This confirms the need to use non coherent light for flat field.
\end{itemize}

Other tests with Gaussian PRF fit (on earlier data) have shown that including pixel offsets does not lead to significant improvement (from $1.9\e{-3}$ to $1.8\e{-3}$ pixel). With the current state of the analysis code, it is only possible to take pixel positions into account with the Gaussian PSF fit (not with the autocorrelation fit), this is the next planned development. Other sources of errors including flat field seem to dominate the errors and have to be dealt with before pixel offsets are taken into account. From Fig. \ref{pixelOffsetsQ1}, which shows an SD of about $2\e{-3}$ for pixel offsets, one expects an error of $3\e{-4}$ on the centroids, which is smaller than the best case residuals ($1.19\e{-3}$).

\begin{figure}[!h]
\begin{center}
\includegraphics[height = 80mm]{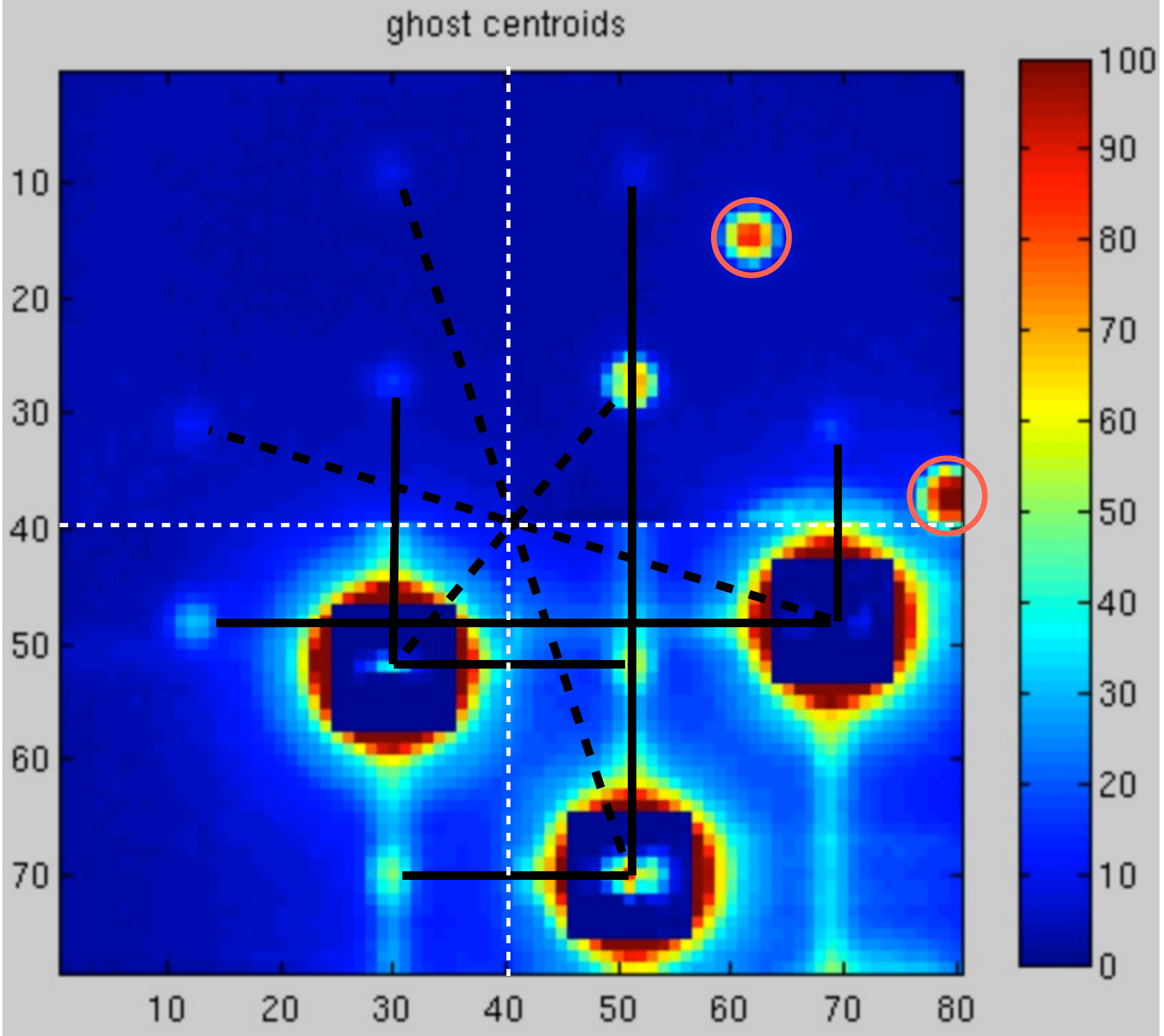}
\caption{\label{ghost_centroids_symmetry}\textbf{Image of the ghost centroids.} There are 3 main stars in the image. It has been processed: the fitting windows (11 by 11 for this test) are replaced by the residuals and the values below 0 and above 100 are truncated. The residuals inside the fitting windows are caused by the centroids motion (this image is extracted from a video). The two brightest ghosts (circled in pink) are defects in the pinhole mask, they are faint but real pseudo stars. They are however co moving the main stars and thus always stays outside of the fitting window. The problematic ghosts are the smaller ones, which are caused by coupling between quadrants during the CCD readout. Their symmetries with respect to the main stars are indicated by black lines (dotted for central symmetry, plain for reflection symmetry). The dotted white line indicates the limits between quadrants. These ghosts can cross the fitting windows when moving the CCD with the translation stage and bias the measurements. Some ghosts are rather faint, they might not be seen on black and white printed paper.}
\end{center}
\end{figure}

One major other identified source of error are the ghost centroids. They are created by a coupling between the quadrants of the CCD. The intensity of the ghost centroids are at a level between 1 to $3\e{-3}$ (depending on the quadrants). Figure \ref{ghost_centroids_symmetry} shows an image of the ghosts.

\section{CONCLUSION}

The NEAT testbed has been assembled and is fully operational, the first tests in vacuum at -10$\dg$C have been made. The ultimate goal is to demonstrate the feasibility of measuring centroids to a precision of $5\e{-6}$ pixel to enable astrometry as a technique for searching Earth-like exoplanets in the habitable zone of nearby stars.

So far, the best measurements have showed an accuracy of $2\e{-4}$ pixel for the position of individual pixels, $3\e{-5}$ for blocs of pixels, $1.2\e{-3}$ for the measure of centroid positions at any location on the CCD and $1.5\e{-5}$ for small perturbations of a few thousandths of a pixel.

Investigation of residuals has shown the presence of ghosts centroids because of coupling between quadrants at a relative level between 1 and $3\e{-3}$, way above the specification needed ($4\e{-6}$) and a strange behavior limited to one CCD quadrant. These two problems, along with the residual stray light and the quality of the flat field have been identified as the most critical issues that must be addressed to improve the performance. Actions are being taken to investigate these problems and make the needed modifications on the testbed and the signal processing methods.


\acknowledgments     

We would like to thank the engineering team at IPAG for their support. We acknowledge the labex OSUG@2020 and CNES for financing the experiment and CNES and Thales Alenia Space for funding the PhD of A.\ Crouzier. At last, we are grateful to Bijan Nemati, Chengxing Zhai and Inseob Hahn for hosting Antoine Crouzier into their team and taking the time to answer many of our questions. 


\bibliography{spie_article_antoine_crouzier_2014}   

\begin{thebibliography}{10}

\bibitem{Malbet11}
{Malbet}, F., {L{\'e}ger}, A., {Shao}, M., {Goullioud}, R., {Lagage}, P.-O.,
  {Brown}, A.~G.~A., {Cara}, C., {Durand}, G., {Eiroa}, C., {Feautrier}, P.,
  {Jakobsson}, B., {Hinglais}, E., {Kaltenegger}, L., {Labadie}, L.,
  {Lagrange}, A.-M., {Laskar}, J., {Liseau}, R., {Lunine}, J., {Maldonado}, J.,
  {Mercier}, M., {Mordasini}, C., {Queloz}, D., {Quirrenbach}, A., {Sozzetti},
  A., {Traub}, W., {Absil}, O., {Alibert}, Y., {Andrei}, A.~H., {Arenou}, F.,
  {Beichman}, C., {Chelli}, A., {Cockell}, C.~S., {Duvert}, G., {Forveille},
  T., {Garcia}, P.~J.~V., {Hobbs}, D., {Krone-Martins}, A., {Lammer}, H.,
  {Meunier}, N., {Minardi}, S., {Moitinho de Almeida}, A., {Rambaux}, N.,
  {Raymond}, S., {R{\"o}ttgering}, H.~J.~A., {Sahlmann}, J., {Schuller}, P.~A.,
  {S{\'e}gransan}, D., {Selsis}, F., {Surdej}, J., {Villaver}, E., {White},
  G.~J., and {Zinnecker}, H., ``High precision astrometry mission for the
  detection and characterization of nearby habitable planetary systems with the
  nearby earth astrometric telescope (neat),'' {\em Experimental Astronomy} ,
  109 (Sep 2011).

\bibitem{Malbet12}
Malbet, F., Goullioud, R., Lagage, P., Léger, A., Shao, M., Crouzier, A., and
  consortium NEAT, ``Neat: a space born astrometric mission for the detection
  and characterization of nearby habitable planetary systems,'' {\em Proc. of
  SPIE}~{\bf 8442} (2012).

\bibitem{Malbet13}
Malbet, F., Crouzier, A., Léger, A., and al., ``Neat: an astrometric mission
  to detect nearby planetary systems down to the earth mass,'' {\em Proc. of
  SPIE}~{\bf 8864} (2013).

\bibitem{Malbet14}
Malbet, F., Crouzier, A., Léger, A., Shao, M., Goullioud, R., and Duigou,
  J.-M.~L., ``Neat: ultra-precise differential astrometry to characterize
  planetary systems with earth-mass exoplanets in the vicinity of our sun,''
  {\em Proc. of SPIE}~{\bf 9143} (2014).

\bibitem{sim_double_blind_test09}
Traub, W., Ford, E., Laughlin, G., Levison, H., Lin, D., Raymond, S., Makarov,
  V., Casertano, S., Fischer, D., Kasdin, J., Muterspaugh, M., Shao, M.,
  Beichman, C., Boss, A., Gould, A., and Marr, J., ``Overview of the sim-rv
  double-blind simulation to detect earths in multi-planet systems,'' in [{\em
  American Astronomical Society Meeting Abstracts
  \#213}{\nolinebreak\hspace{0.1em}]},  {\em Bulletin of the American
  Astronomical Society} {\bf 41},  \#300.01 (Jan 2009).

\bibitem{neat_number_of_measurements11}
Malbet, F. and Léger, A., ``Neat document: number of visits needed per
  target,'' Private communication (2011).

\bibitem{neat_error_budget11}
Goullioud, R., ``Neat error budget,'' Private communication (2011).

\bibitem{Crouzier12}
Crouzier, A., Malbet, F., Preis, O., Henault, F., Kern, P., Martin, G.,
  Feautrier, P., Cara, C., Lagage, P.~O., Léger, A., LeDuigou, J.~M., Shao.,
  M., and Goullioud, R., ``An experimental testbed for neat to demonstrate
  micro-pixel accuracy,'' {\em Proc. of SPIE}~{\bf 8445} (2012).

\bibitem{Crouzier13}
Crouzier, A., Malbet, F., Preis, O., Henault, F., Kern, P., Martin, G.,
  Feautrier, P., Cara, C., Lagage, P.~O., Léger, A., LeDuigou, J.~M., Shao.,
  M., and Goullioud, R., ``First experimental results of very high accuracy
  centroiding measurements for the neat astrometric mission,'' {\em Proc. of
  SPIE}~{\bf 8864} (2013).

\bibitem{Henault14}
Hénault, F., Crouzier, A., A., F.~M., and consortium NEAT, ``Neat breadboard
  system analysis and performance models,'' {\em Proc. of SPIE}~{\bf 9150}
  (2014).

\end{thebibliography}
\bibliographystyle{spiebib}   

\end{document}